\documentclass[12pt,a4paper,english,superscriptaddress,aps,nofootinbib]{revtex4}
\usepackage{amsmath,amssymb,graphicx}
\makeatletter
\usepackage{babel}
\usepackage[active]{srcltx}
\usepackage{graphicx,color}
\usepackage{changebar}
\usepackage{hyperref}
\hypersetup{
	colorlinks=true,
	urlcolor= blue,
	citecolor=blue,
	linkcolor= blue,
	bookmarks=true,
	bookmarksopen=false,}

\usepackage[T1]{fontenc}
\usepackage{ae}
\usepackage[ansinew]{inputenc}
\usepackage{amsmath}
\usepackage{braket}
\usepackage{mathtools}
\usepackage{slashed}
\usepackage{empheq}
\usepackage{tikz}
\usepackage{multirow}
\usetikzlibrary{decorations.pathmorphing}
\usepackage[caption = false]{subfig}

\usepackage{graphicx}
\usepackage{amsfonts}
\usepackage{amsmath}
\newcommand{\be}{\begin{equation}}
	\newcommand{\ee}{\end{equation}}
\newcommand{\bea}{\begin{eqnarray}}
	\newcommand{\eea}{\end{eqnarray}}

\begin{document}
	\title{Exponentially generalized vortex}
	
\author{F. C. E. Lima}
\email[]{E-mail: cleiton.estevao@fisica.ufc.br}
\affiliation{Universidade Federal do Cear\'{a} (UFC), Departamento do F\'{i}sica - Campus do Pici, Fortaleza, CE, C. P. 6030, 60455-760, Brazil.}

\author{C. A. S. Almeida}
\email[]{E-mail: carlos@fisica.ufc.br}
\affiliation{Universidade Federal do Cear\'{a} (UFC), Departamento do F\'{i}sica - Campus do Pici, Fortaleza, CE, C. P. 6030, 60455-760, Brazil.}	

\begin{abstract}
\vspace{0.5cm}
\begin{center}
\large{Abstract}
\end{center}
In this work, we propose an exponentially generalized Abelian model. We investigated the presence of vortex structures in models coupled to Maxwell and Chern-Simons fields. We chose to investigate the dynamics of the complex scalar field in models coupled separately to the Maxwell term and the Chern-Simons term. For this, we analyze the Bogomol'nyi equations in both cases to describe the static field configurations. An interesting result appears when we note that scalar field solutions generate degenerate minimum energy configurations by a factor of $\nu^{2}$ in Maxwell's case. On the other hand, in the case of Chern-Simons, the solutions in this sector are degenerate by a factor of $\kappa\nu^{2}/a_{s}$. Finally, we solve the Bogomol'nyi equations numerically and discuss our results.
\end{abstract}

\maketitle
\thispagestyle{empty}
\newpage

\section{Introduction}

Vortices are field configurations in $(2+1)D$ obtained due to the interaction of the gauge field with the scalar field. These structures have a finite energy configurations and for that, it is required a local $U(1)$ symmetry of the scalar field and consequently the introduction of a minimally coupled gauge field \cite{nielsen}. The great interest in the study of vortex solutions is due to the important applications of topological defects in high energy physics \cite{Bunkov}, where the structures emerge as static solutions of field equations \cite{manton,weinberg}. In this context, these structures can play an important role in understanding the early universe and in cosmic evolution \cite{Vilenkin}. In condensed-matter physics, these structures appear when we study the behavior of superconductors and its magnetic properties \cite{singh,abrikosov,golod}.

Understanding topological defects allows to predict properties of materials \cite{abrikosov,chernodub} or cosmological objects \cite{Polchinski,Hanany,jackiw}. Currently, we have a vast and interesting scenario of vortex investigations in the literature \cite{babichev,lima,lima1,flood,forgacs,1,2,3,LA12}. Some examples of topological vortices studies appear in investigations of pipelike field configurations in models with global symmetries $SU(2)$ and $U(1)$ local. These models lead to extended topological structures with longitudinal currents  and involve two scalar fields \cite{chernodub}. Also, in the so called Witten vortices, a symmetry $U(1)\times U(1)$ with two components \cite{witten}, involves two scalar fields. It is interesting to mention that it is always possible to modify the models in various ways and to investigate specific characteristics of their topological structures \cite{lee,lima}.

As discussed in Ref. \cite{shifman}, vortices can also be investigated in models with expanded symmetries to describe various fields and to increase the degrees of freedom of the theory. Field theories that discuss symmetries $U(1)\times U(1)$ can be useful to include the so-called hidden sector, which is interesting, since these theories can provide an explanation for the origin of dark matter \cite{long,long1}. Basically, the hidden sector can be coupled to the visible sector through the scalar fields or through the coupling with a gauge field \cite{silveira,dienes}.

The generalized models have become promising to describe some properties and new classes of topological solutions as presented in refs. \cite{lima,lima1}. Interesting solutions arise when generalizations are made in the models, a clear example is the discussion of compact solutions describing the dynamics of the Higgs field \cite{casana}. This result is stimulating, since it was initially addressed in the study of fluid dynamics described by the well-known KdV equation \cite{KdV}.  

Theory-$k$ including generalized kinematic terms, has been a subject of great interest in recent years \cite{casana}. The existence of vortex solutions in a generalized Abelian model was studied in ref. \cite{ghosh}; while the generalized Chern-Simons model was studied in \cite{lee}. Some generalizations are realized by introduction of some generalized parameter or function in kinetic terms. These generalized theories first emerged as effective cosmological models to describe inflationary evolution \cite{picon}. Soon, studies in tachyon matter problem \cite{sen}, gravitational wave \cite{muknakov},  and ghost condensates \cite{arkani} appeared.  Other studies show that it is possible in generalized theories to obtain topological solitons \cite{babichev1,adam,adam1}.

In the early 1930s, Hideki Yukawa showed that interaction or exchange of a massive scalar field with a boson field, also massive, generates a potential \cite{brown}. In this context, considering that the approximate range of the nuclear force was known, the Yukawa equation could be used to predict the  mass of the particle in the force field, even before it was discovered \cite{brown,yukawa}. In the case of nuclear force, this mass was predicted to be about 200 times the mass of the electron, and this was later considered to be a prediction of the existence of the pion, before it was detected in 1947 \cite{griffiths}. Nowadays, the so called Yukawa potential (exponential potential)  is applied in several braches of physics. For instance, an exponential potential mediating the interaction of dark matter particles could explain the observed cores in dwarf galaxies \cite{loeb}.

In this letter, we will present exponentially generalized Abelian topological models. For this, our objective is to investigate topological structures in theories exponentially generalized under the influence of the Maxwell and Chern-Simons fields. In Maxwell's model this new coupling is introduced by a exponential generalization term in the kinetic gauge field, while in the Chern-Simons model we insert the exponential generalization in the scalar kinetic term. Then, we study the stationary solutions using the BPS approach. Finally, we present the numerical results obtained and conduct a brief discussion on the models studied.

\section{The generalized Maxwell model}

Let us start working in $(2+1)$ flat spacetime dimensions with the Lagrangian density
\begin{align}\label{LLa}
    \mathcal{L}=-\frac{1}{4}G(\vert\phi\vert)F_{\mu\nu}F^{\mu\nu}+H(\vert\phi\vert)\vert D_{\mu}\phi\vert^{2}-V(\vert\phi\vert).
\end{align}

This Lagrangian with a function $G(|\phi|)$ generalizing the kinetic term for the gauge field was introduced for the first time by Lee and Nam \cite{lee} in order to find soliton solutions of an Abelian Higgs model. Lima et al. \cite{LPA} used it to find vortices for different profiles of the generalizing function. Models with generalized dynamics, i. e., with a non-canonical dynamic, have been investigated in Refs. \cite{LA,LPA1}. In Ref. \cite{DBazeia}, Bazeia et. al. consider a similar Lagrangian density and shows that magnetic vortices are self-dual in the absence of shearing forces (stressless condition). In other words, self-dual structures are obtained if $T_{ij}=0$. Other generalized Maxwell-Higgs models can be seen in Refs. \cite{DBazeia1,DBazeia2,DBazeia3}.

Here the scalar field $\phi$ is complex and $|\phi|^{2}=\bar{\phi}\cdot\phi$. The notation $\bar{\phi}$ representing the complex conjugate of $\phi$. $A_{\mu}$ is the Abelian gauge field, $F_{\mu\nu}=\partial_{\mu}A_{\nu}-\partial_{\nu}A_{\mu}$ is the electromagnetic tensor. The potential is given by $V(|\phi|)$ and the metric is $\eta_{\mu\nu}=\text{diag}(+, -, -)$. 

The covariant derivative is defined as
\begin{equation}
D_{\mu}\phi=\partial_{\mu}\phi+ieA_{\mu}\phi.
\end{equation}

We are interested in the study of vortices, these structures arise in our generalized model due to the gauge group $U(1)$ and the Higgs mechanics in (2+1)-dimensional space-time. Thus, all fields that describe the vortices of our generalized model must be invariant under the following transformations:
\begin{align}\label{trans1}
     \phi'(r)\to \text{e}^{i\alpha(r)}\phi(r), \hspace{0.5cm} \text{and} \hspace{0.5cm}  A_{\mu}'(r)\to A_{\mu}(r)+\frac{1}{e}\partial_\mu \alpha(r).
\end{align}
By direct calculation, we have that $D_{\mu}\phi'=$e$^{i\alpha(r)} D_\mu\phi$, so that $\vert D_{\mu}\phi '\vert^2=\vert D_{\mu}\phi'\vert^2$. The invariance of the electromagnetic field under the transformation (\ref{trans1}) is immediately seen. Therefore, in the generalized case the vortices are structures invariant over the transformations (\ref{trans1}), and the only possible profile for the generalization functions is if it is a function purely dependent on the modulus of the matter field, i. e. $G(\vert\phi\vert)$ and $H(\vert\phi\vert)$. More details on the choice of our generalization are discussed in the next section.

The equations of motion associated with Lagrangian density (\ref{LLa}) are
\begin{align} \label{em}
    &H(\vert\phi\vert)\cdot D_{\mu}D^{\mu}\phi+\frac{\phi}{2\vert\phi\vert}\bigg(\frac{G_{\vert\phi\vert}}{4}F_{\mu\nu}F^{\mu\nu}+V_{\vert\phi\vert}\bigg)=0,\\ \label{em1}
    &\partial_\mu (G F^{\mu\nu})=J^{\nu},
\end{align}
where the current is $J_\mu=iHe (\overline{\phi}D_\mu\phi-\phi\overline{D_\mu \phi})$, $G_{\vert\phi\vert}=\partial G/\partial\vert\phi\vert$, and $V_{\vert\phi\vert}=\partial V/\partial\vert\phi\vert$. 

Spacetime translation symmetry yields the following energy-momentum tensor
\begin{align}\label{Emomentum}
    T_{\mu\nu}=GF_{\mu\rho}F^{\rho}\,_{\nu}+H\overline{D_{\mu}\phi}D_\nu \phi+H D_{\mu}\phi\overline{D_\nu \phi}-\eta_{\mu\nu}\mathcal{L}.
\end{align}

For the study of these structures, we should choose $A_0=0$. Choosing $A_0=0$, the Gauss' law is preserves (i.e., the $\nu=0$ component of Eq. (\ref{em1})) in the theory. To investigate static and rotationally symmetric configurations, we consider the ansatz:
\begin{align}
\label{ansatz}
\phi=g(r)\text{e}^{in\theta}, \hspace{0.5cm} \text{and} \hspace{0.5cm} \vec{A}=-\frac{1}{er}[a(r)-n]\hat{\theta},
\end{align}
where $n \, \in \, \mathbb{Z}$ and depicts the vorticity (or winding number).

Considering the ansatz (\ref{ansatz}) and remembering that
\begin{equation}
\label{magnetic1}
\vec{B}=\vec{\nabla}\times\vec{A},
\end{equation}
arrives at
\begin{equation}
\label{magnetic}
B\equiv ||\vec{B}||=-\frac{a'(r)}{er}=-F_{12},
\end{equation}
where $a'(r)$ denotes the derivative of $a$ with respect to the radial coordinate $r$. The magnetic flux $\Phi_{B}$ of the vortex is given by 
\begin{align}
\label{flux}
\Phi_{B}=-\int_{0}^{2\pi}\int_{0}^{\infty} F^{12} r\, dr\, d\theta=-\frac{2\pi}{e}[a(\infty)-a(0)].
\end{align}

If we assume that the asymptotic behavior for fields are
\begin{equation} \label{condition}
g(0)=0, \hspace{0.5cm} g(\infty)=\nu, \hspace{0.5cm} a(0)=n, \hspace{0.5cm} a(\infty)=0,
\end{equation}
the vortex will have a quantized magnetic flux given by
\begin{equation}
\Phi_{B}=\frac{2\pi n}{e}.
\end{equation}

To investigate static field configurations it is necessary to write the $T_{00}$ component of the energy-momentum tensor in terms of the field variables (\ref{ansatz}), i. e.,
\begin{align}
    \mathcal{E}=G(g)\frac{a'(r)^2}{2e^2r^2}+H(g)g'(r)^2+H(g)\frac{a(r)^2g(r)^2}{r^2}+V(g).
\end{align}

In order to investigate structures with BPS property, let us organize the energy functional as follows:
\begin{align}\nonumber
    \mathcal{E}=& \frac{G(g)}{2}\bigg[\frac{a'(r)}{er}\pm\frac{e(\nu^2-g^2)}{G(g)}\bigg]+H(g)\bigg(g'(r)\mp\frac{a(r)g(r)}{r}\bigg)+V(g)-\frac{e^2(\nu^2-g^2)^2}{2G(g)}\\
    \mp& \frac{1}{r}[a(r)(\nu^2-g(r))]'\pm\frac{2ag(r)g'(r)}{r}(H(g)-1).
\end{align}

For the model to have BPS property, we choose
\begin{align}\label{VV}
    V(g)=\frac{e^2(\nu^2-g^2)^2}{2G(g)}, \hspace{1cm} \text{and} \hspace{1cm} H=1,
\end{align}
so that the energy density is rewritten as follows:
\begin{align}
    \mathcal{E}=& \int_{0}^{2\pi}\int_{0}^{\infty}\frac{G(g)}{2}\bigg[\frac{a'(r)}{er}\pm\frac{e(\nu^2-g^2)}{G(g)}\bigg]\, rdrd\theta+\int_{0}^{2\pi}\int_{0}^{\infty}\bigg(g'(r)\mp\frac{a(r)g(r)}{r}\bigg)\, rdrd\theta+ E_{BPS},
\end{align}
where
\begin{align}\label{Em1}
    E_{BPS}= 2\pi\vert n\vert\nu^2.
\end{align}

Note that the energy is limited by $E_{BPS}$, i. e. $E\geq E_{BPS}$. Therefore, if the solutions obey the first order equations:
 \begin{align}
     g'(r)=\pm \frac{a(r)g(r)}{r}, \hspace{1cm} \text{and} \hspace{1cm} \frac{a'(r)}{er}=\mp\frac{e(\nu^2-g(r)^2)}{G(g)},
 \end{align}
the BPS limit is reached with the energy saturation, i. e.,  $E=E_{BPS}$. Note that for the model to admit BPS properties $H=1$, i. e. the topological structures at the BPS limit must be governed by a usual canonical dynamic.

\subsection{Exponential electric permeability}

We highlight that the dielectric function profile is not a standard choice in the literature. For each objective, it is necessary to consider a specific form for the dielectric function (see for example ref. \cite{LPA}). From previous works \cite{LPA}, we know that a generalized model by non-polynomial functions can change the shape of the topological structure, and make it more energetic, which can be useful for its handling and experimental detection. The centerpiece of our study are  the electromagnetic vortex structures with an exponential dielectric permeability. Indeed, we believe that this discussion is presented for the first time in this work. The non-polynomial dielectric permeability function change some characteristics of the topological structures \cite{LPA}. This motivates us to understand how an exponential medium will change our electromagnetic vortices. Furthermore, functions with exponential profiles have been used in cosmological scenarios, for the study of inflationary models, see f. e. Ref. \cite{FABrito}. Not too far, when studying topological structures through Hirota's bilinear formulation, the addition of an exponential function allows detecting the emergence of lump-kink \cite{Liu} structures. Studies of solitons in exponential networks also have been used to understand the properties this topological structures \cite{Yuan}.

In this work, let us  consider an exponential profile for the permeability function. We expect that the permeability function of the exponential type changes the topological profile of the theory and intensifies the magnetic flux and energy of the structure. An important point is that spontaneous symmetry breaking must also be preserved in theory for the description of solitons to be possible. Therefore, taking into account that the model must be gauge invariant (i. e. invariant under the transformations (\ref{trans1}) and the generalization (which relates to the potential at the BPS limit, i. e., $V(\vert\phi\vert)\propto G(\vert\phi\vert) ^{-1}$) must preserve the spontaneous symmetry breaking, we consider that the simplest form for electrical permeability with an exponential profile is
\begin{align}\label{perm}
    G(\vert\phi\vert)=-\frac{a_s}{\vert\phi\vert}\text{e}^{-\lambda\vert\phi\vert},
\end{align}
where $a_s$ and $\lambda$ are adjustable parameters that are responsible for the emergence of topological structures. In a similar theory, for example, Yukawa's theory, the parameter $a_s$ is related to the force acting on the particle. In this context, this parameter is related to field compactification. This is a property of  finite wavelength solitons \cite{hyman}, which was addressed by many authors to study cosmic strings \cite{nielsen}, and compact vortices \cite{lima2,LA}.

At the energy saturation limit, the equations of motion of the model are rewritten as follows:
\begin{align}
\label{VMaxwell}
    g'(r)=\pm\frac{a(r)g(r)}{r} \hspace{1cm} \text{and} \hspace{1cm}     \frac{a'(r)}{er}=\pm\frac{e[\nu^{2}-g(r)^{2}]g(r)}{a_s\text{e}^{-\lambda g(r)}}.
\end{align}

To reach the BPS property of the model, the interaction (\ref{VV}) of the generalized exponential case assumed the profile
\begin{align}\label{VV1}
    V(g)=-\frac{e^2}{2a_s}\frac{[\nu^2-g(r)^2]^2g(r)}{\text{e}^{-\lambda g(r)}}.
\end{align}

\begin{figure}[ht!]
\centering 
\includegraphics[height=6cm,width=7.5cm]{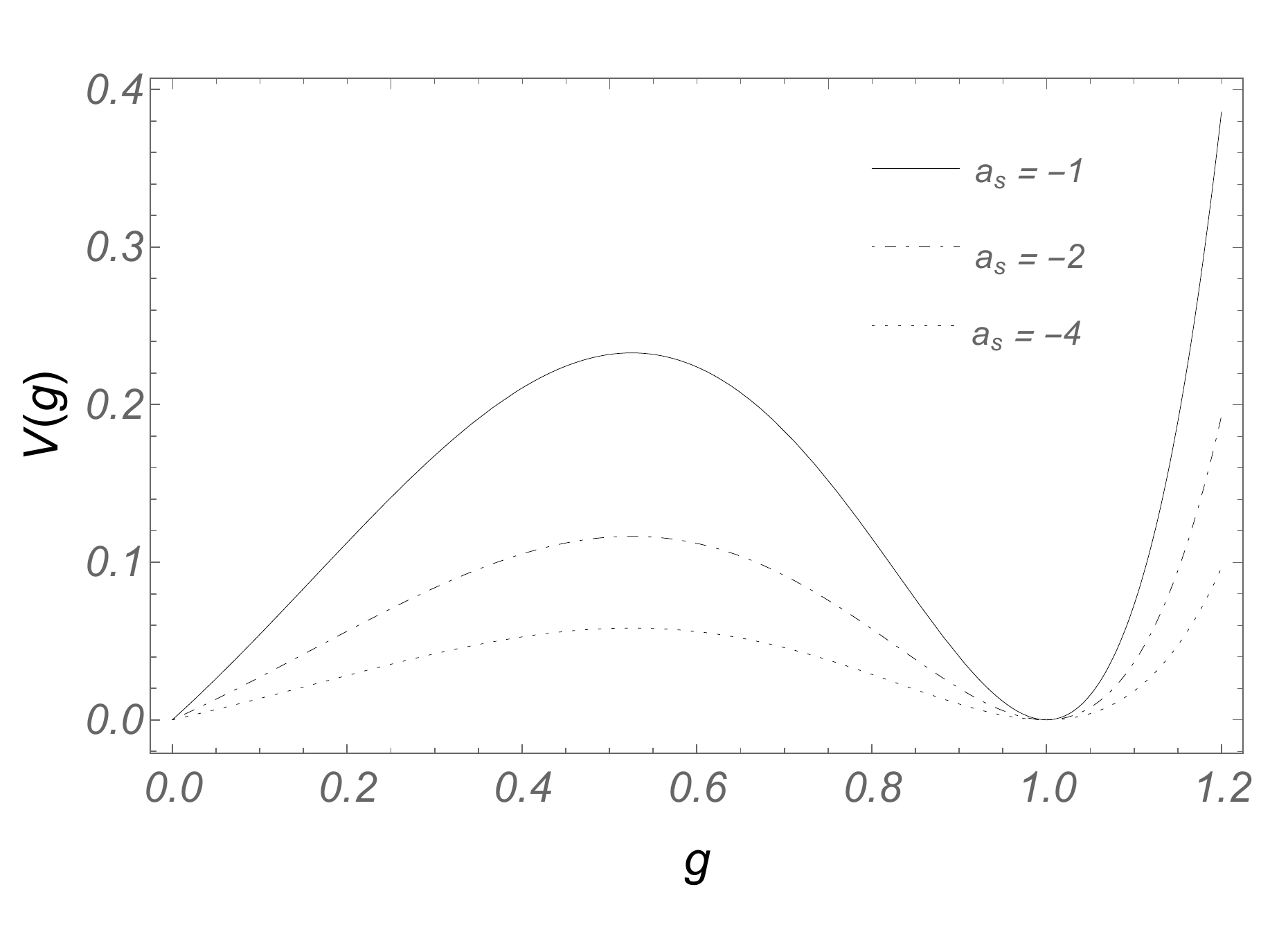}
\vspace{-20pt}
    \caption{Behavior of the potential for several values of $a_s$. Similar behavior is obtained for $\nu=\lambda=1$.}
    \label{PY}
\end{figure}

The plot of the potential (\ref{VV1}) that spontaneously breaks the symmetry and allows the appearance of topological structures is shown in Fig. \ref{PY}.
 
\begin{figure}[ht!]
\centering
\includegraphics[height=6cm,width=7.5cm]{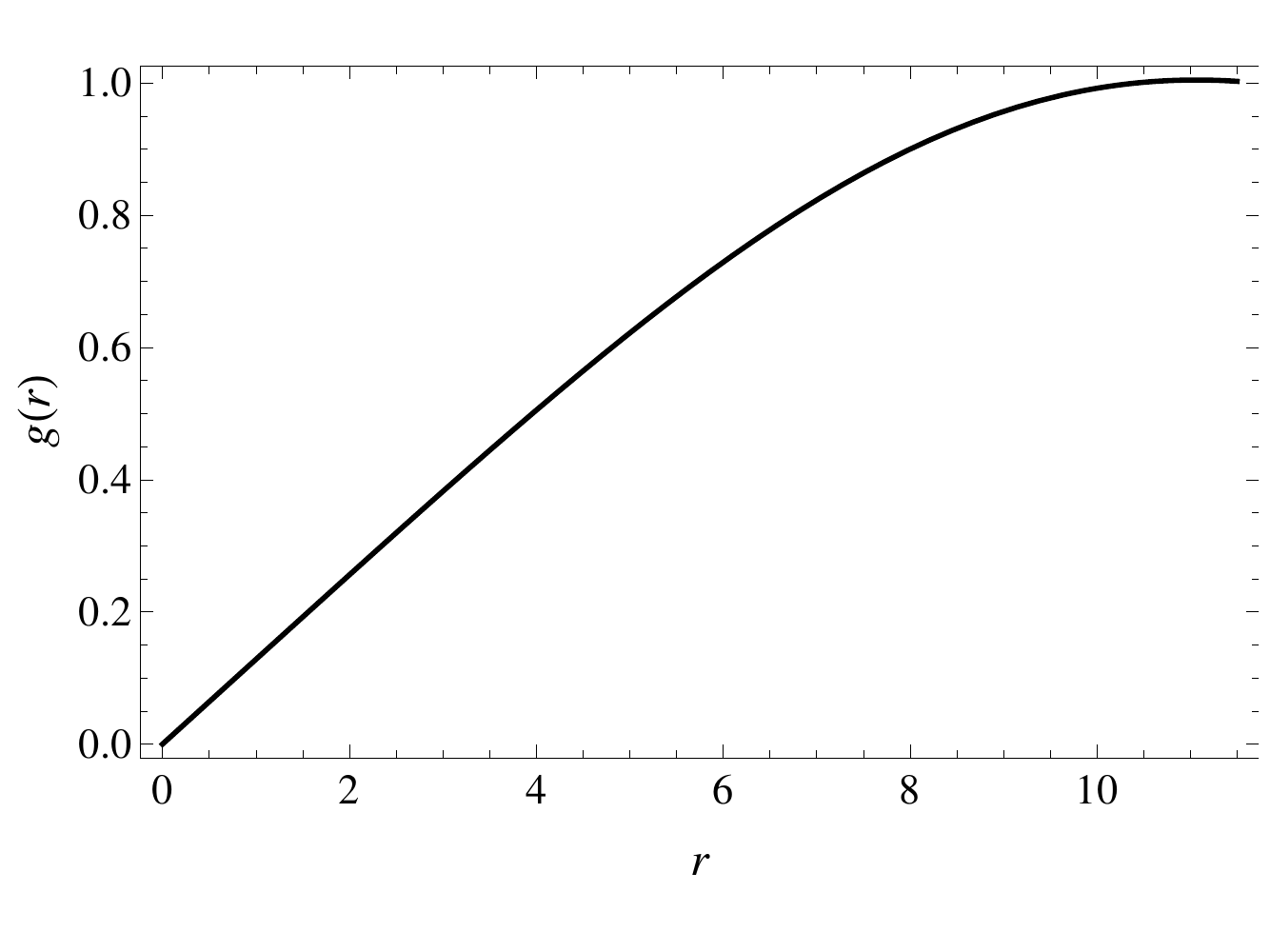}
\includegraphics[height=6cm,width=7.5cm]{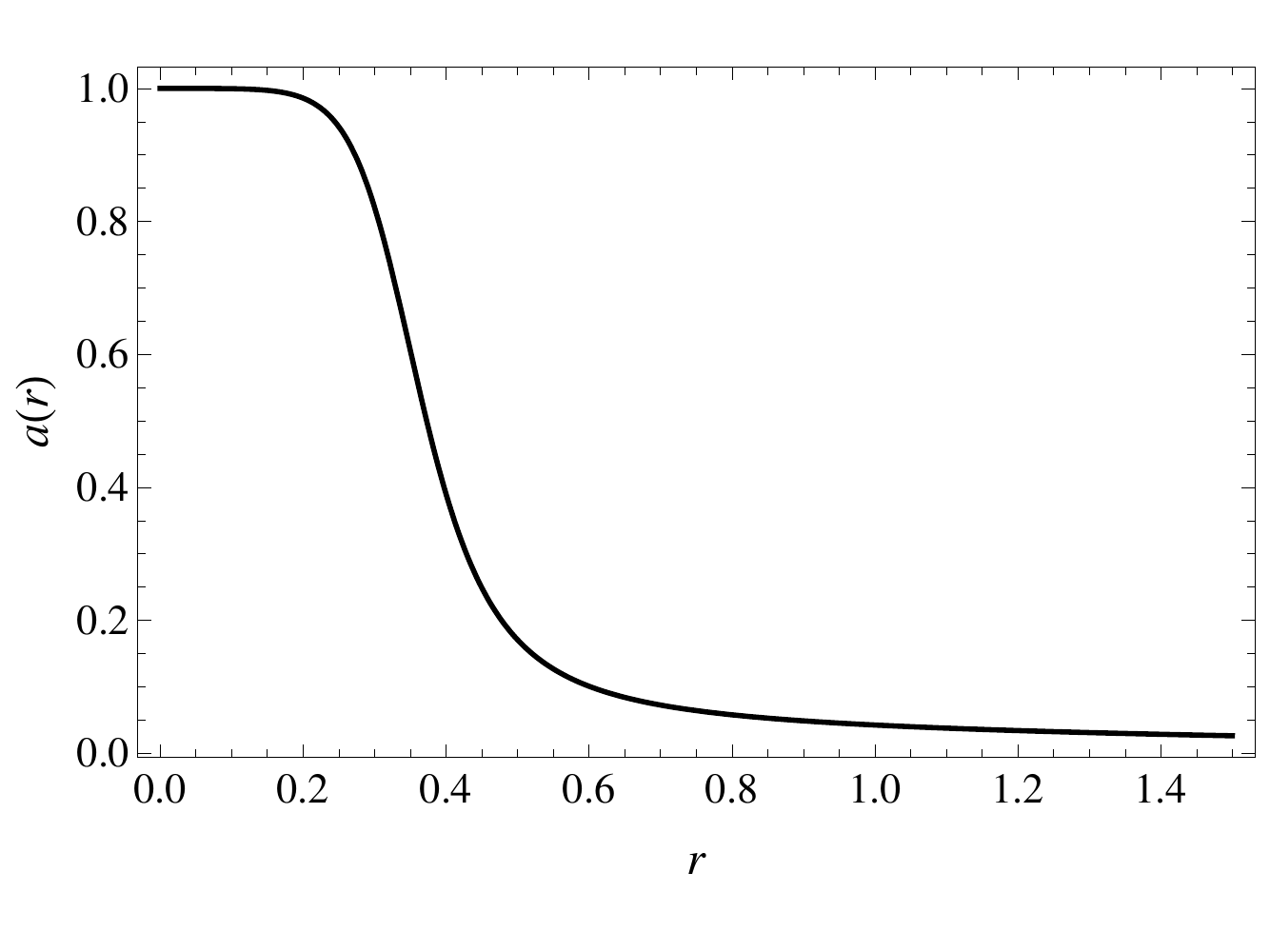}
\vspace{-20pt}
\caption{Numerical solution of the scalar field $g(r)$ and the gauge field $a(r)$ with vorticity $n=1$, $\nu=1$, $\lambda=0.7$ and $a_{s}=-1$.}
\label{fig1}
\end{figure}

At this point, we will turn our attention to the numerical study of Bogomol'nyi equations to describe the dynamics of the complex scalar field and the gauge field. For this, we start from the equations (\ref{VMaxwell}). We assume that the variable field has a behavior near the origin proportional to $r^{n}$, where $n$ is vorticity of the model. Then, by means of a numerical interpolation to the topological limits presented in Eq. (\ref{condition}), we obtain the topological solution presented in Fig. (\ref{fig1}).

If we have a matter field solutions from eq. (\ref{VMaxwell}), we can describe the gauge field $a(r)$ of the structure. The gauge field is  showed in Fig.(\ref{fig1}).

\begin{figure}[ht!]
\centering
\includegraphics[height=6cm,width=7.5cm]{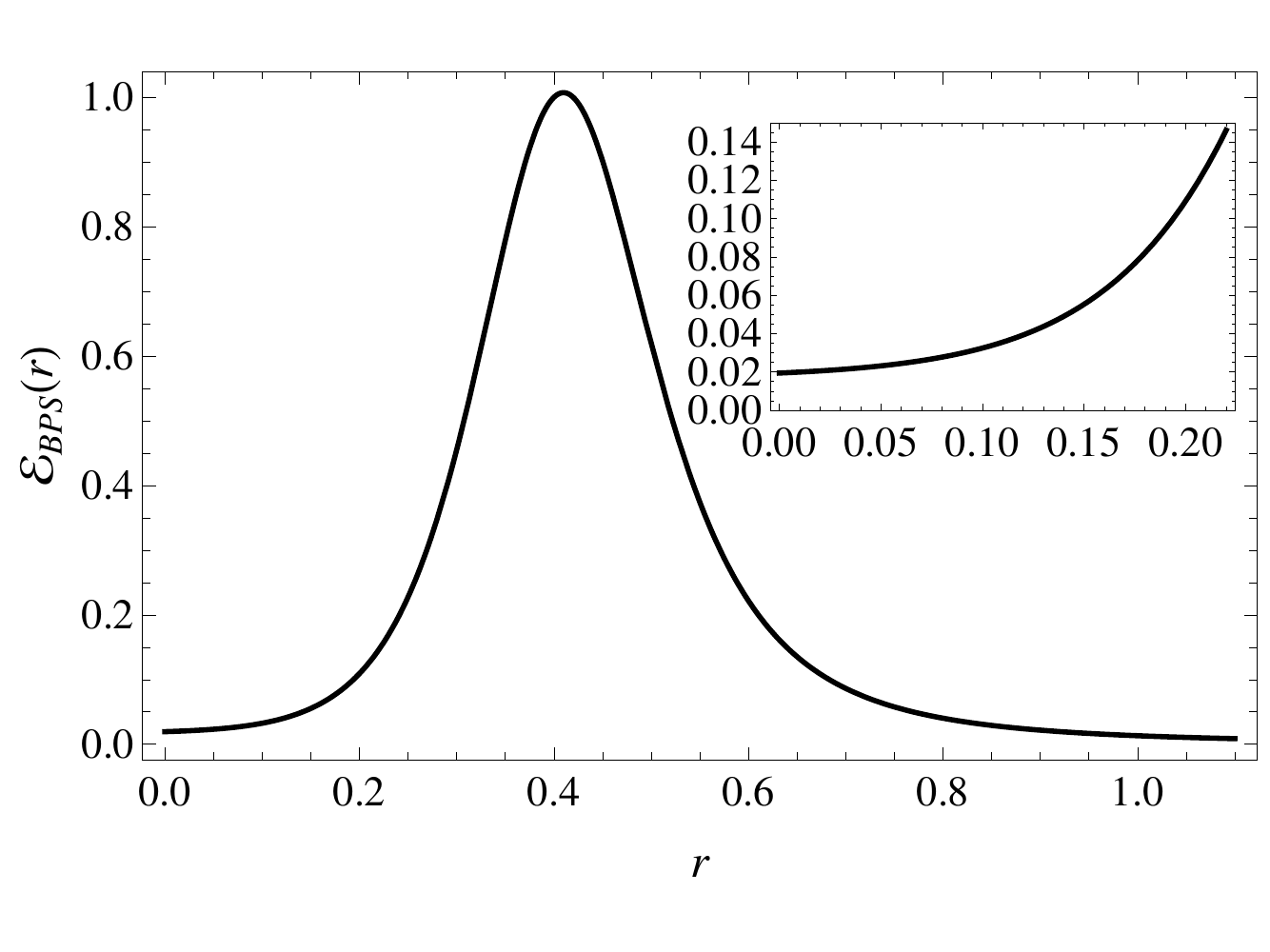}
\includegraphics[height=6cm,width=7.5cm]{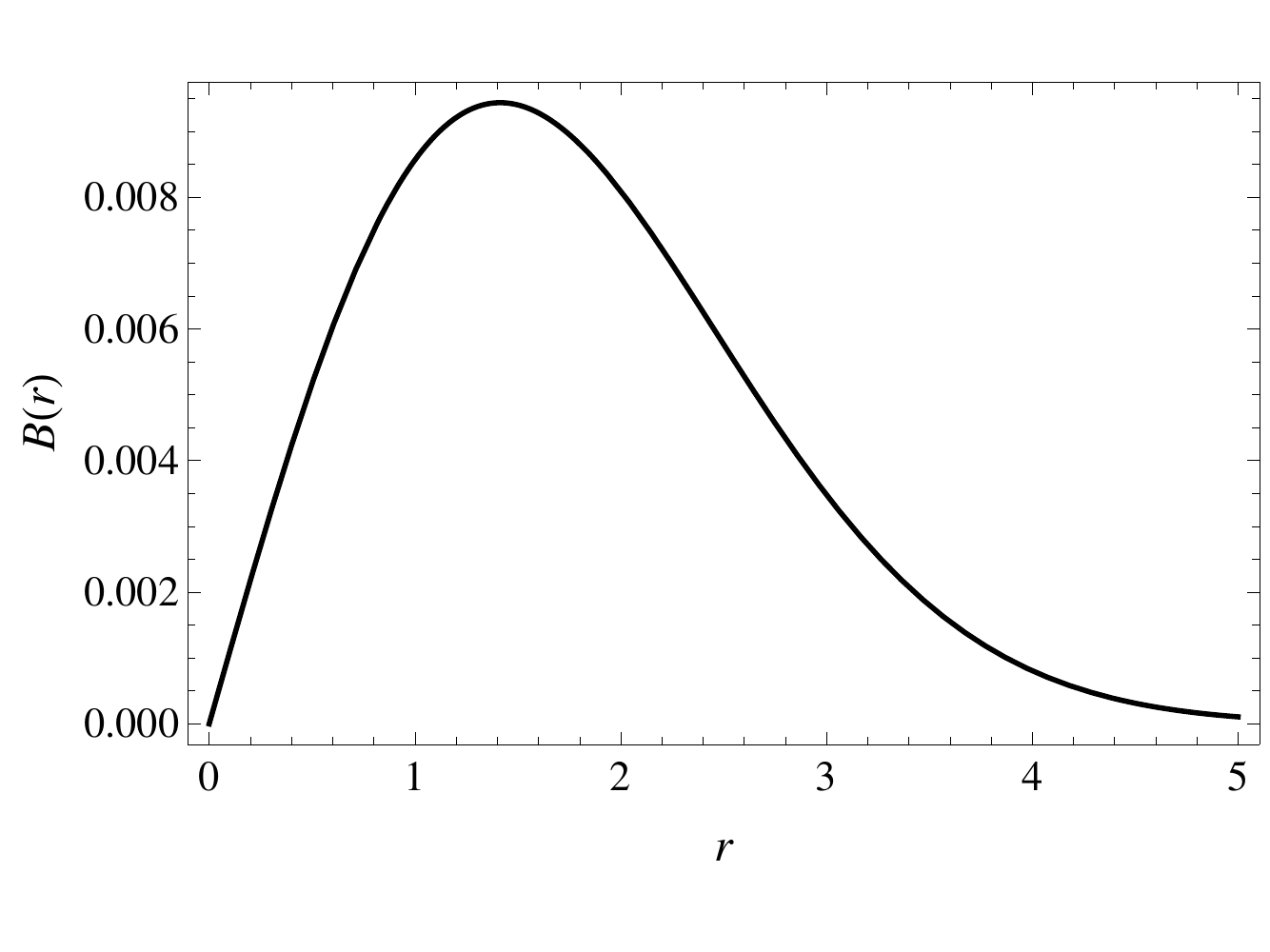}\\
\vspace{-20pt}
\begin{center}
\hspace{0.5cm}    (a) \hspace{7cm} (b)
\end{center}
\vspace{-20pt}
\caption{(a) BPS energy density of the structures. (b) Behavior of the magnetic field of the vortex.}
\label{fig3}
\end{figure}

We now turn our attention to the numerical study of the BPS energy density for the field configurations showed in the previous figure. With this in mind, remembering that the model's energy is described by equation (\ref{Em1}), the energy density has the shape showed in Fig.(\ref{fig3}(a)).

On the other hand, from the equation (\ref{magnetic}), we investigate the behavior of the magnetic field (Fig. \ref{fig3}(b)) that generates the vortex structures.

As expected, the solutions lead us to topological field configurations that have finite and positive-defined energy. We noticed that the magnetic field has a shape similar to a Gaussian curve. Also, it is easy to notice that when the scalar field tends to the value of the vacuum state of the model the magnetic flux of the vortex will be zero. As a direct consequence, at this point, the BPS energy density will tend to zero, i. e., $\mathcal{E}_{BPS}\rightarrow 0$ when $g(r)\rightarrow\nu$.

\section{The generalized Chern-Simons model}

For the study of vortex structures in the Chern-Simons model, we consider the generalized Lagrangian density in the form
\begin{equation}
    \label{LChernSimons}
    \mathcal{L}=-\frac{a_{s}\text{e}^{-\lambda |\phi|}}{|\phi|}\overline{D_{\mu}\phi}D^{\mu}\phi+\frac{\kappa}{4}\varepsilon^{\alpha\beta\gamma}A_{\alpha}F_{\beta\gamma}-V(|\phi|).
\end{equation}

In fact, to study the topological vortices of the generalized theory, the term of $G(\vert\phi\vert)$ must be coupled with the kinetic term to ensure that the theory is gauge-invariant (i. e. the theory is invariant under the transformations (\ref{trans1}). This is due to the fact that the Chern-Simons field does not support the inclusion of a coupling with the exponential function (or any function $G(\vert\phi\vert)$ ) as this would lead to a theory that is not invariant under the gauge transformation. In fact, this type of model has its relevance in the study of soliton bag models of quarks and gluons and supersymmetric theories. 

For this case, let us consider that the gauge field is $A^{\mu}=(A_{0}, \vec{A})$. Therefore, the electric and magnetic fields are
\begin{align}
   \label{ElectricFieldCS}
   \vec{E}=-\dot{A}^{i}-\partial_{i}A^{0} \hspace{0.25cm} \text{and}  \hspace{0.25cm} \vec{B}=-F^{12}.
\end{align}

By varying the action associated with Lagrangian density (\ref{LChernSimons}), we obtain that the equations of motion for the scalar field is
\begin{align}\nonumber
    D_{\mu}\bigg(\frac{a_{s}\text{e}^{-\lambda\vert\phi\vert}D^{\mu}\phi}{\vert\phi\vert}\bigg)=&-\frac{\phi}{2|\phi|}\bigg[\frac{a_{s}\text{e}^{-\lambda |\phi|}}{|\phi|}\bigg(\lambda+\frac{1}{|\phi|}\bigg)\overline{D_{\mu}\phi}D^{\mu}\phi-V(|\phi|)\bigg].
\end{align}

For the gauge field, we have
\begin{equation}
    \frac{\kappa}{2}\varepsilon^{\lambda\mu\nu}F_{\mu\nu}=J^{\lambda},
\end{equation}
where
\begin{equation}
    J_{\mu}=-iea_{s}\frac{\text{e}^{-\lambda |\phi|}}{|\phi|}(\overline{\phi}D_{\mu}\phi-\phi\overline{D_{\mu}\phi}).
\end{equation}

To investigate the finite energy field configurations, we need the energy-momentum tensor, namely,
\begin{equation}
    T_{\mu\nu}=-\frac{a_{s}\text{e}^{-\lambda |\phi|}}{|\phi|}(\overline{D_{\mu}\phi}D_{\nu}\phi+\overline{D_{\nu}\phi}D_{\mu}\phi)-\eta_{\mu\nu}\mathcal{L}.
\end{equation}

In order to investigate the vortex structures, we consider the previous ansatz for the scalar field and for the gauge field (\ref{ansatz}). We assume that $A_{0}=A_{0}(r)$. Again, the magnetic field is given by
\begin{equation}
\label{MagneticCFieldS}
    B=-\frac{a'(r)}{er}.
\end{equation}

This magnetic field leads us to vortices of quantized magnetic flux as discussed in the generalized Maxwell model. 

The vortices carry an electrical charge given by
\begin{equation}
    Q_{e}=2\pi\int\, rdr\, J^{0}=-\kappa\Phi\rightarrow\Phi=\frac{Q_{e}}{\kappa}.
\end{equation}

These are vortices that are electrically charged and with non-quantized flux.

Writing the equations of motion of the model in terms of the variable field, we obtain
\begin{align}\nonumber
-\frac{1}{r}\bigg(\frac{a_{s}\text{e}^{-\lambda g(r)}}{g(r)}r g'(r)\bigg)-&a_{s}\text{e}^{-\lambda g(r)}\bigg(e^{2}A_{0}^{2}-\frac{a(r)}{r^{2}}\bigg)+\frac{1}{2}\bigg[\frac{a_{s}\text{e}^{-\lambda g(r)}}{g(r)}\bigg(\lambda+\frac{1}{g(r)}\bigg)\times \\ 
&\bigg(e^{2}g(r)^{2}A_{0}^{2}-g'(r)^{2}-\frac{a(r)^{2}g(r)^{2}}{r^{2}}\bigg)-V_{g}\bigg]=0,
\end{align}
\begin{align}
    \frac{a'(r)}{r}-\frac{2e^{3}g(r)^{2}A_{0}}{\kappa}\frac{a_{s}\text{e}^{-\lambda g(r)}}{g(r)}=0,
\end{align}
and
\begin{align}
    \label{ElectricFieldCS1}
    A_{0}-\frac{2ea_{s}a(r)g(r)\text{e}^{-\lambda g(r)}}{\kappa r}=0.
\end{align}

Conveniently, the energy density of the model can be written as

\begin{align}
\label{ChernSimonsEnergy}\nonumber
    \mathcal{E}=&\bigg(\frac{\kappa a'(r)}{2e^2 r\sqrt{b_{s}g(r)\text{e}^{-\lambda g(r)}}}\pm\sqrt{V(g)}\bigg)^{2}+\frac{b_{s}\text{e}^{-\lambda g(r)}}{g(r)}\bigg(g'(r)\mp\frac{a(r)g(r)}{r}\bigg)^{2}\pm\frac{a(r)g'(r)}{r}\times\\  &\bigg(\frac{\kappa}{e^{2}}\frac{d}{dg}\sqrt{\frac{V(g)}{b_{s}g(r)\text{e}^{-\lambda g(r)}}}+2b_{s}\text{e}^{-\lambda g(r)} \bigg)\mp\frac{1}{r}\bigg(\frac{\kappa}{e^{2}}a(r)\sqrt{\frac{V(g)}{b_{s}g(r)\text{e}^{-\lambda g(r)}}}\bigg)',
\end{align}
where we assume that $b_{s}=-a_{s}$.

Note that, if we consider
\begin{equation}
    \frac{d}{dg}\sqrt{\frac{V(g)}{b_{s}g(r)\text{e}^{-\lambda g(r)}}}=-2b_{s}\text{e}^{-\lambda g(r)},
\end{equation}
the energy density that describes the field configuration is reduced to
\begin{align} \nonumber
\mathcal{E}=&\bigg(\frac{\kappa a'(r)}{2e^2 r\sqrt{b_{s}g(r)\text{e}^{-\lambda g(r)}}}\pm\sqrt{V(g)}\bigg)^{2}+\frac{b_{s}\text{e}^{-\lambda g(r)}}{g(r)}\bigg(g'(r)\mp\frac{a(r)g(r)}{r}\bigg)^{2}\mp\frac{1}{r}\bigg(\frac{\kappa}{e^{2}}a(r)\times\\
&\sqrt{\frac{V(g)}{b_{s}g(r)\text{e}^{-\lambda g(r)}}}\bigg)'.
\end{align}

Therefore, at the bound of energy saturation (BPS bound), the equations of motion are reduced to
\begin{align}
\label{bps3}
    &g'(r)=\pm\frac{a(r)g(r)}{r},\\ 
    &\kappa a'(r)=\mp 2 e^2 r \sqrt{b_{s}g(r)\text{e}^{-\lambda g(r)}V(g)},
\end{align}

and the BPS energy is 
\begin{align}
    \label{EnergyBPSCS}
    E_{BPS}&=
    \mp 2\pi\int_{0}^{\infty}\,\bigg(\frac{\kappa}{e^{2}}a(r)\sqrt{\frac{V(g)}{b_{s}g(r)\text{e}^{-\lambda g(r)}}}\bigg)'\, dr.
\end{align}

Again, considering the expression (\ref{VV1}), we conclude that the BPS energy of the model is
\begin{equation}
    E_{BPS}=\frac{\sqrt{2}\pi\kappa\nu^{2}|n|}{ea_{s}}.
\end{equation}

We clearly perceive that the topological solutions of the Chern-Simons model are degenerate in this sector by a factor of $\kappa\nu^2/a_{s}$.

We rewrite the BPS Eqs. (\ref{bps3}), considering Eq. (\ref{VV1}), which leads us to 
\begin{align}
    \label{BpsCS}
    &g'(r)=\pm\frac{a(r)g(r)}{r},\\
    &\kappa a'(r)=\mp\sqrt{2}e^{3}rg(r)^{3/2}[\nu^{2}-g(r)^{2}].
\end{align}


\subsection{Numerical result}

In order to get a more comprehensive analysis, it is interesting now to accomplish numerical solutions of the BPS equations. The numerical solutions of the generalized Chern-Simons model describe the topological solutions of the scalar field $g(r)$ and the gauge field $a(r)$. To start, we consider, without loss of generality, that the field $g(r)$ assumes a behavior near the origin of the type
\begin{equation}
    g(r)=\sum_{n}c_{n}r^{n}.
\end{equation}

\begin{figure}[ht!]
\centering
\includegraphics[height=6cm,width=7.5cm]{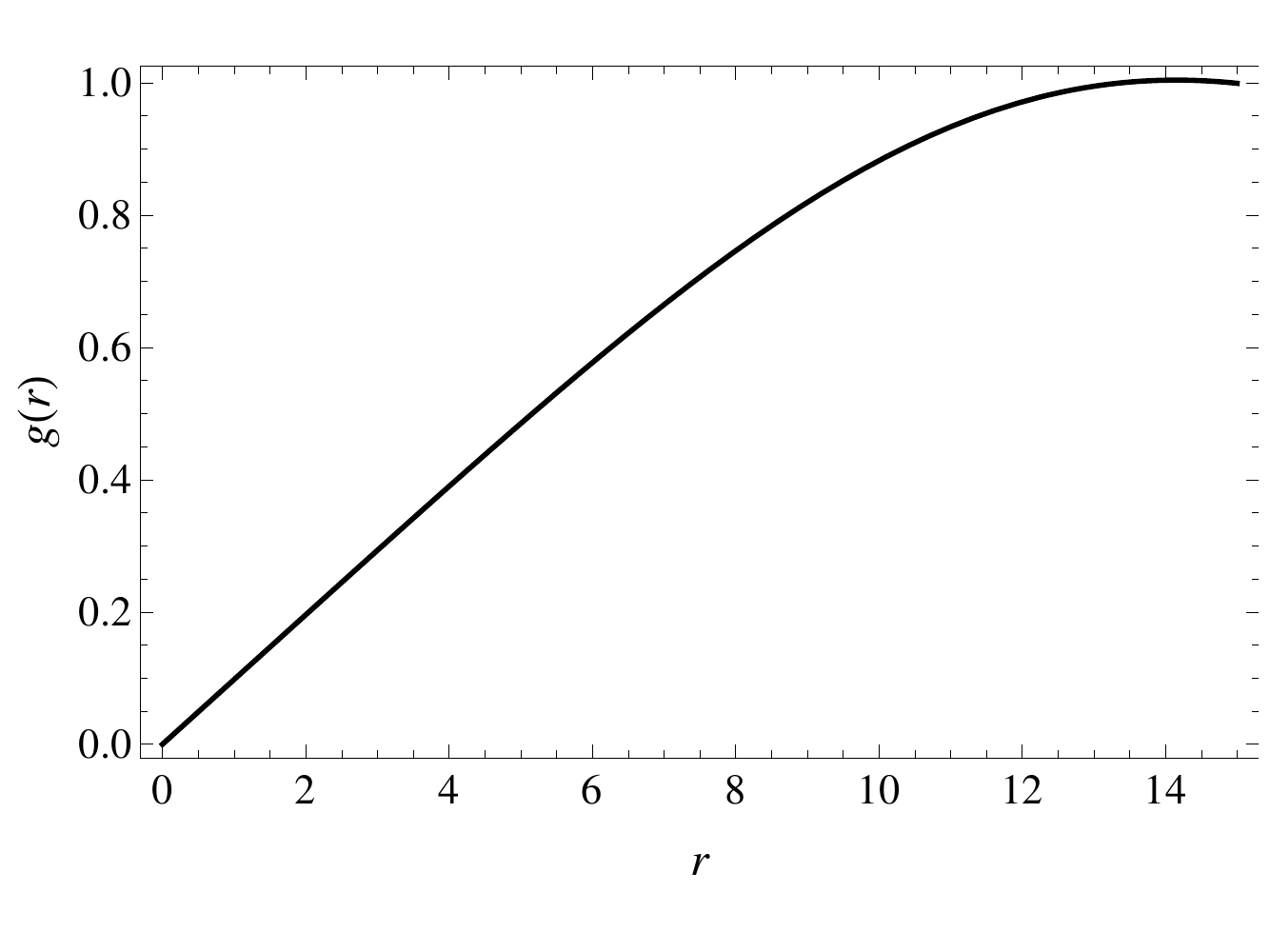}
\includegraphics[height=6cm,width=7.5cm]{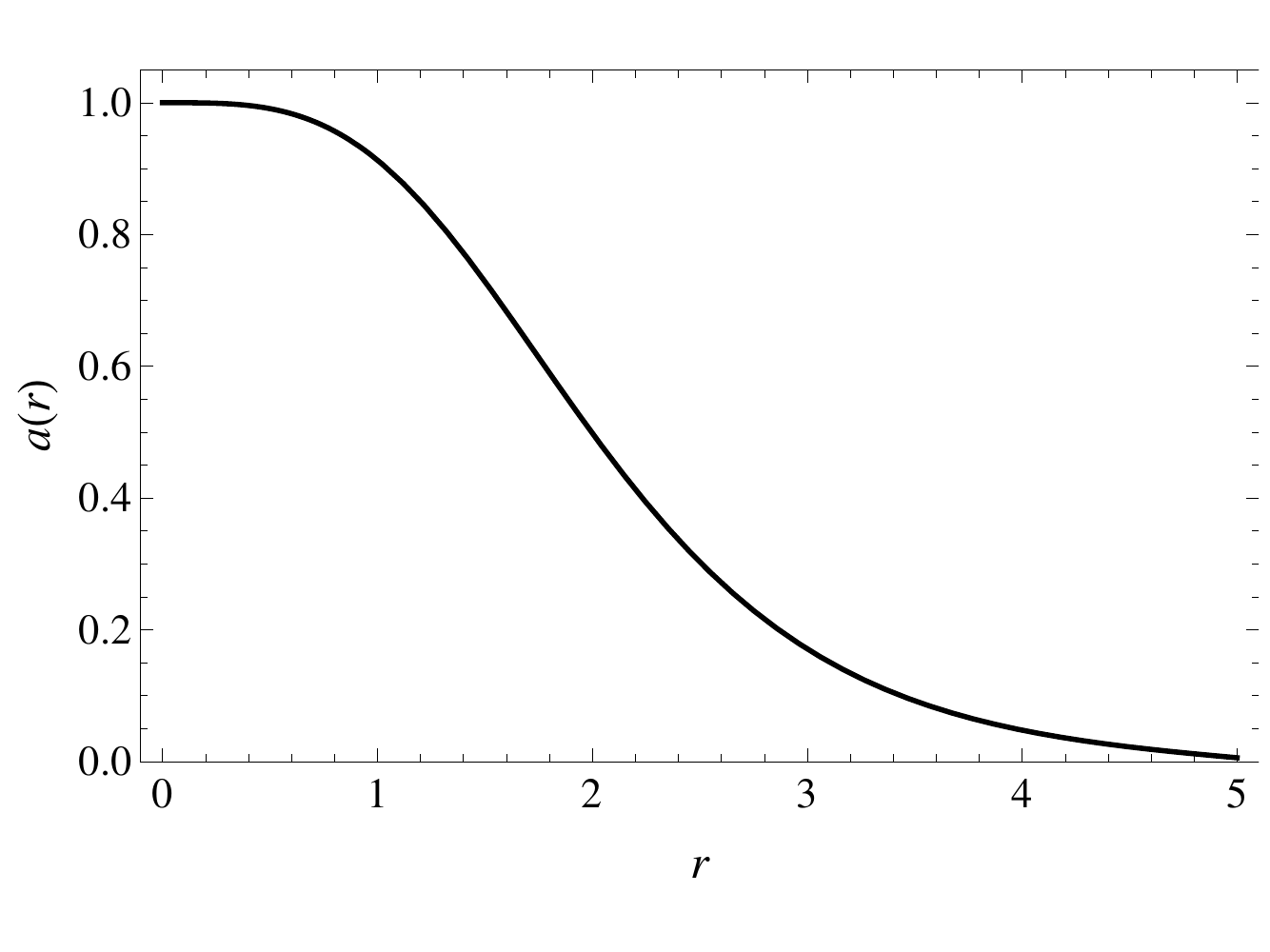}
\vspace{-20pt}
\caption{Numerical solution of the scalar field $g(r)$ and the gauge field $a(r)$ with vorticity $n=1$, $\nu=1$, $\lambda=0.7$ and $a_{s}=-1$.}
\label{fig4}
\end{figure}

In this way,  we obtain the topological solutions for the fields (See Fig. (\ref{fig4})), decoupling the Eqs. (\ref{BpsCS}) and applying the interpolating method with the asymptotic boundary conditions of the Eq. (\ref{condition}).

\begin{figure}[ht!]
\centering
\includegraphics[height=6cm,width=7.5cm]{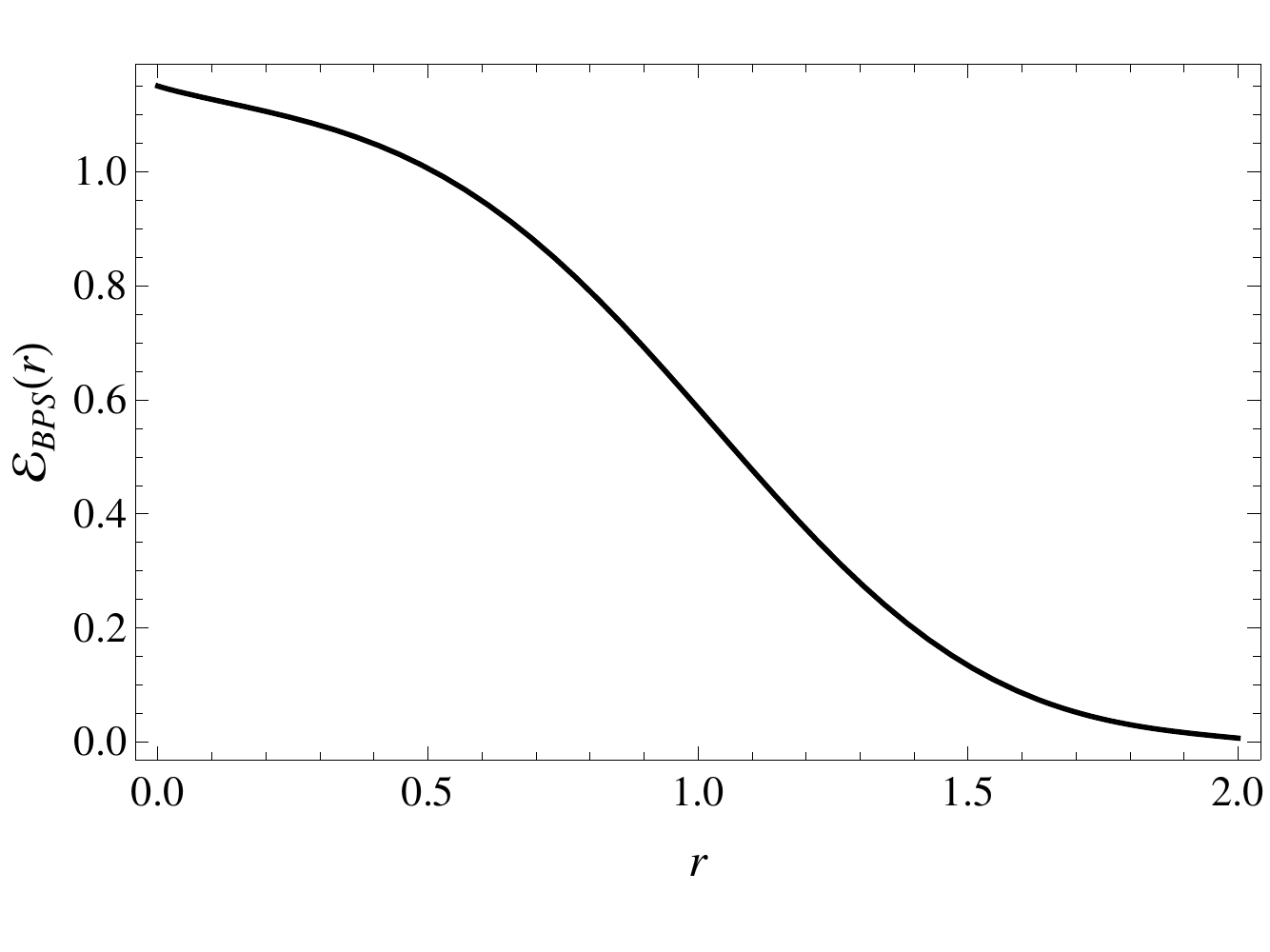}
\includegraphics[height=6cm,width=7.5cm]{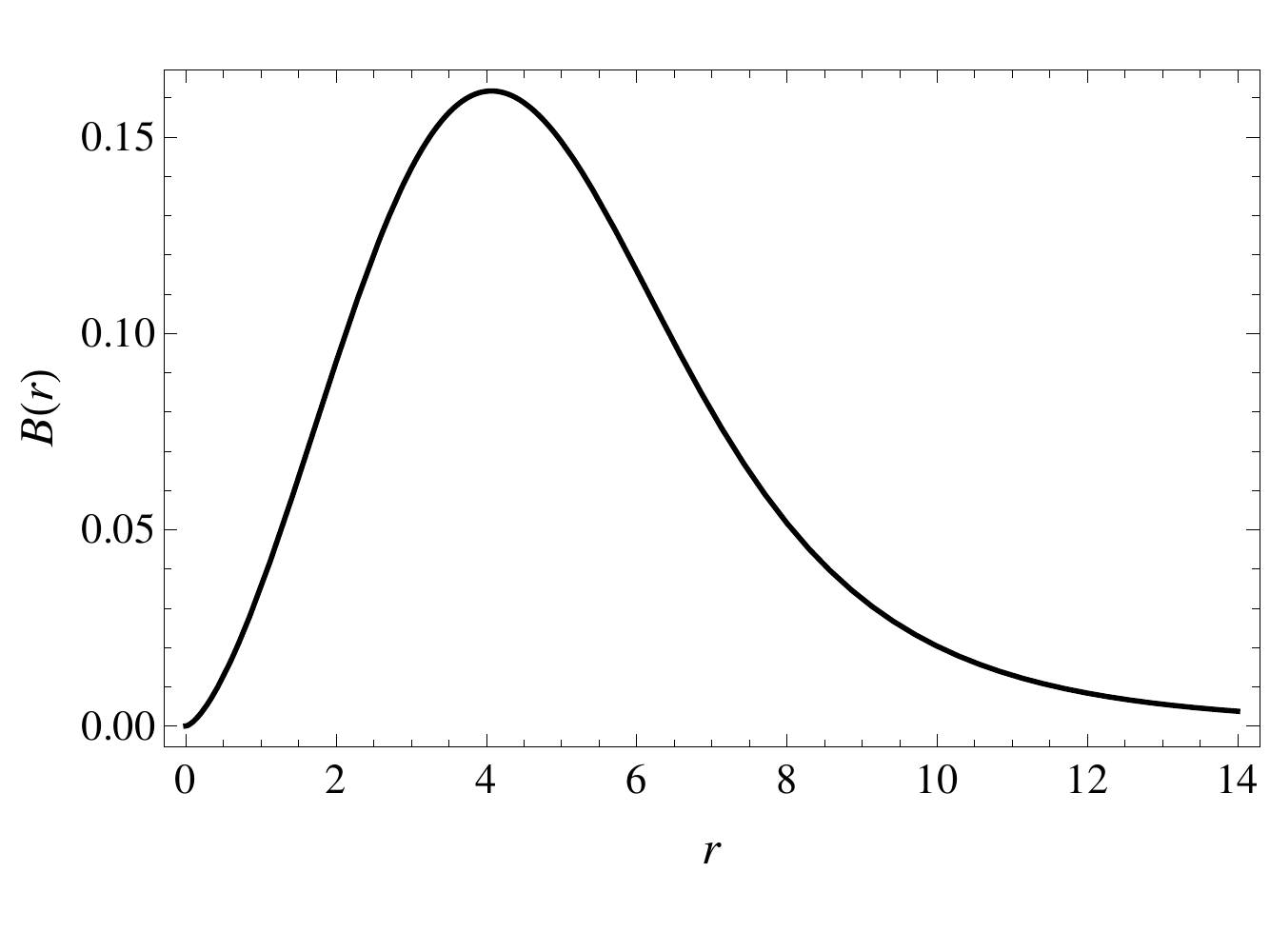}\\
\vspace{-20pt}
\begin{center}
\hspace{0.5cm}    (a) \hspace{7cm} (b)
\end{center}
\includegraphics[height=6cm,width=7.5cm]{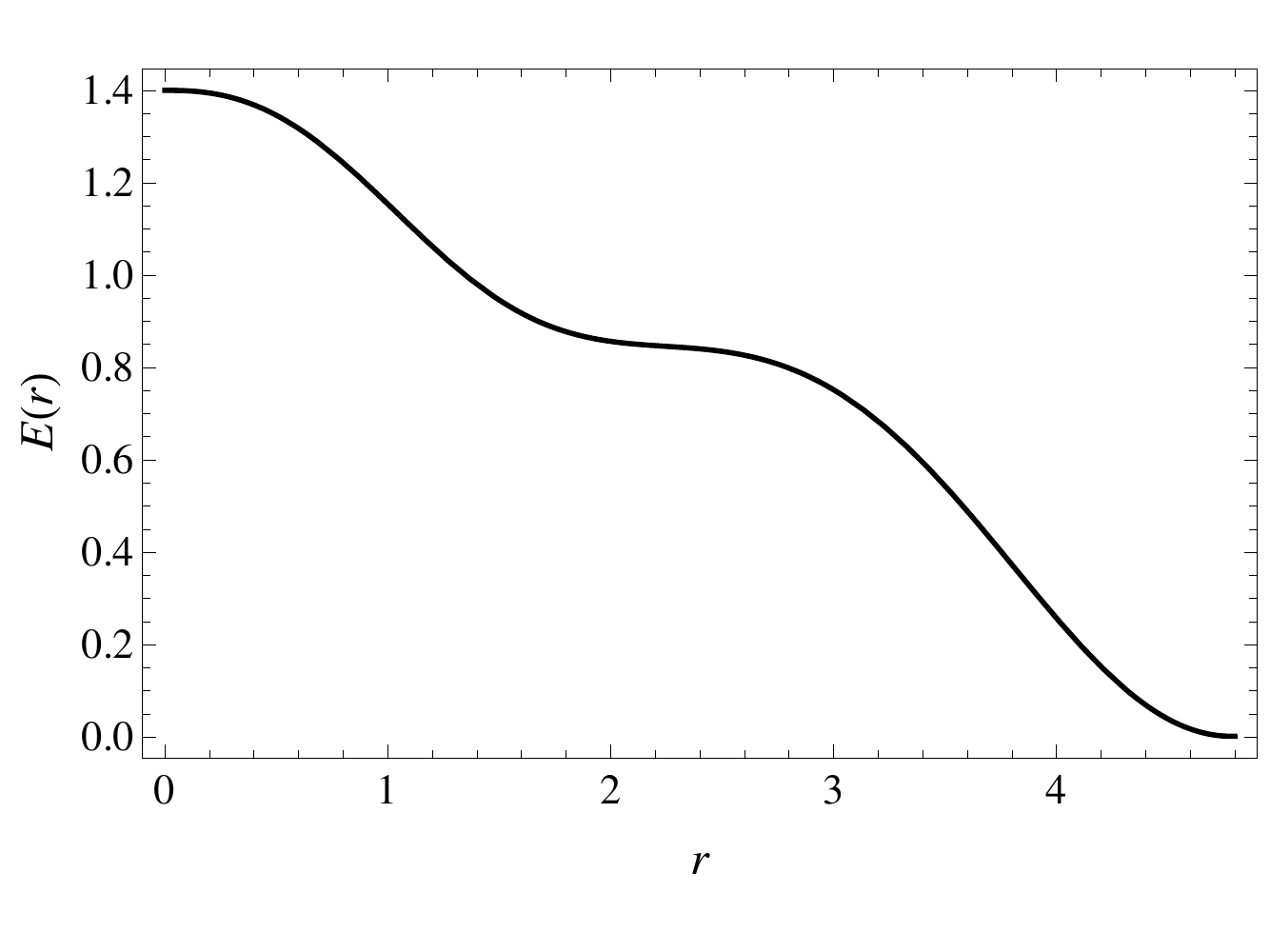}
\vspace{-20pt}
\begin{center}
 (c)
\end{center}
\vspace{-20pt}
\caption{(a) BPS energy density of the vortex. (b) Behavior of the magnetic $B(r)$. (c) Electric field.}
\label{fig6}
\end{figure}

With the field solutions described by eqs. (\ref{BpsCS}) and represented by Fig. (\ref{fig4}), we now use the expression (\ref{EnergyBPSCS}) and describe the BPS energy density of the generalized Chern-Simons model showed in Fig. (\ref{fig6}(a)).

Finally, the behavior of the magnetic field (Fig. \ref{fig6}(b)) and the electric field (Fig. \ref{fig6}(c)) can be analyzed using eqs. (\ref{MagneticCFieldS}), (\ref{ElectricFieldCS}) and (\ref{ElectricFieldCS1}).

\section{Final remarks}

In this letter, our main objective was to study the exponentially generalized topological vortex structures of the Maxwell and Chern-Simons abelian models. A particularly interesting result emerges when analyzing the generalized Maxwell model. That is, for obtaining configurations of the self-dual field, we have that in the limit of energy saturation, the kinetic term have the usual form, i. e. $\vert D_{\mu}\phi\vert^2$. This result shows one equivalence of the structures of the generalized canonical model with stressless condition (results shown in Ref. \cite{DBazeia}) and the generalized Maxwell model with usual canonical dynamics. In this form, for the study of the generalized model, our problem is restricted to the behavior of the dielectric permeability function.

An interesting feature of the Maxwell model with exponential permeability is that it differs significantly from the nonpolynomial structures studied in ref. \cite{LPA}, this is because the Maxwell structures in our model generate a ringlike profile when we study the energy densities and magnetic field of the structure in the plane. A curious behavior in the exponentially generalized vortex of Maxwell's theory is that it is less energetic in the core of the structure and near to the point $r=0.4$ it has maximum energy. This is seen as a consequence of the profile of the magnetic field of the structure, as it has an almost Gaussian profile off-center from the origin.

To broaden the understanding of these structures it is convenient to compare our result with a similar model and without generalization. A model that meets this criterion is the discussed model Lee and Nam \cite{lee}. Comparing our results with Ref. \cite{lee}, we noticed that the generalized model maintains the guarantee of the existence of topological solitons. However, in our model (generalized model), the presence of non-topological structures was not detected when the electrical permeability function admits an exponential profile. This absence of non-topological structures seems to be a consequence of the dielectric function profile (see f. e. Ref. \cite{lee}).

Meanwhile, upon studying the generalized Chern-Simons model, the generalization term must be coupled to the kinetic term of the theory for the model to be gauged invariant. A consequence of the generalization of the model is coupled to the kinetic term is that for topological structures to have the BPS property, they must have a flux with contributions of magnetic field and electric field. As a result, the magnetic flux will not be quantized. We observe that most of the contribution of the electromagnetic flux of the vortex is due to the contribution of the electric field of the structure. A consequence of this is seen in the energy density of the vortex since the electric field strength near $r=0$ overlap the magnetic field making the BPS energy density of the structure assume a profile similar to the curve sech$^2(r)$. Finally, it is important to mention that the generalization in both models seems to behave by adjusting the intensity of the electromagnetic fields of the structures, making the flux more (or less) energetic.

It is important to highlight that the results described in this letter, can encourage further investigations into the extensions of Maxwell and Chern-Simons models. An immediate extension is the Skyrmions models with other generalization profiles. A natural line of investigation that emerges from our results is the search for topological structures governed by non-Abelian fields. Indeed, these structures may be important for describing holographic superconductors and insulators, as suggested in Refs. \cite{Herzog,Mefford}. The results of generalized topological vortices with a dielectric permeability function, are also useful for understanding electromagnetic metamaterials with negative refractive indices. These materials are described by negative magnetic permeability that depends on a frequency range \cite{Shelby,Rama}. The adaptation of the theoretical description of these materials in a field theory comes through the introduction of the generalization function, which in principle can have an exponential profile.

A continuation of this study could be to consider the extensions of these models in curved space and investigate the influence of topological structures on the geometry of spacetime. We hope to carry out this study in future works.


\section*{Acknowledgments}
We are grateful to the Coordena\c{c}\~{a}o de Aperfei\c{c}oamento do Pessoal de N\'{\i}vel Superior (Grants CAPES - FCEL), and Conselho Nacional de Desenvolvimento Cient\'{\i}fico e Tecnol\'{o}gico (CNPq), grant n$\textsuperscript{\underline{\scriptsize o}}$ 308638/2015-8 (CASA), for financial support. The authors thank the anonymous referee for important criticisms and suggestions.

\end{document}